\documentclass[i2pt]{iopart}

\usepackage{graphicx}
\usepackage{bm}
\usepackage[T1]{fontenc}
\usepackage{txfonts}

\begin{document}

\title{The effect of dipole-dipole interactions between atoms in an active medium}

\author{V.~V.~Berezovsky$^1,^3$, L.~I.~Men'shikov$^1,^2$, S.~\"Oberg$^3$ and C.~D.~Latham$^3$}
\address{$^1$Pomor State University, Lomonosov st., 4, Arkhangelsk, 163002, Russia}
\address{$^2$Russian Research Center ``Kurchatov Institute'', Kurchatov sq., 1, Moscow, 123182, Russia}
\address{$^3$Department of Mathematics, Lule{\aa} University of Technology, \mbox{SE-97187 Lule{\aa}}, Sweden}
\ead{vladimir@sm.luth.se}

\begin{abstract}
Based on the results of numerical modeling, it is shown that dipole-dipole
interactions among atoms in the active medium influences strongly the character
of the associated superradiation.  The main effect is to make the nuclear
subsystem behave chaotically.  Its strength increases with the atom density,
and leads to the suppression of distant collective correlations and
superradiation.  Near correlations between the atoms are established, causing a
confinement effect: a shielding of radiation in the active medium.
\end{abstract}

\maketitle

\section{\label{sec:level1}Introduction and method}
Superradiation (SR) is the cooperative radiation arising in a medium that
contains a population inversion of excited states.  Originally this effect has
been stated for purely quantum systems: i.e.\ two-level
atoms~\cite{dicke-PR-93-99}.  Experiments have confirmed this
prediction~\cite{skribanowitz-PRL-30-309}.  Later work established that this
phenomenon also occurs in classical systems \cite{gap2,gaponov-SR-10-794},
and that the phasing effect---the spontaneous origin and strengthening of
correlations of originally independent subsystems---underlies it.  In the
quantum case, these are correlations among phases of electronic states of atoms
undergoing radiative transitions; while in the classical regime correlations
among phases of oscillations and directions of the electric dipole moments of
atoms occur.  A full account of the influence on SR of the dipole-dipole
interactions among atoms remains incomplete (see
referenses~\cite{gross-PREP-93-301,stenholm-PREP-6-1,men}).
 
The SR theory has been developed in several directions. There exist
complementary to each other Schr\"odinger, Heisenberg and semiclassical
approaches. Each of them is applicable to a special area of values of
the system parameters.  The common methodological lack of these approaches
is that the phasing mechanism remains off screen.  The mechanism of
the transition from casual to a phased state possesses certain
spatial, time and statistical behaviors  and its nature is not fully
clear. The quantum mechanical problem of SR is rather complicated, for
example, within the Heisenberg approach it requires to solve a system
of nonlinear operational equations. Approximations which are used to
simplify this systems have  limited and often unclear area of 
applicability. Classical model of superradiation  (CMS), where atoms
are substituted by the classical Lorenz oscillators and the electromagnetic
field is described by the classical Maxwell equations, allows to answer
many difficult questions, in particular, the phasing mechanism. Therefore classical and quantum approaches
complement each other. Moreover, radiation produced by pure classical
system such as electrons revolved in magnetic field, electron clouds
created in wigglers, cathode-ray lamps for microwaves, etc. is also
SR.   

Let us consider only classical systems.  First, phasing leads to the ordering of
phases of atoms.  Second, according to Earnshaw's theorem \cite{siv,str}, a
system of point dipoles cannot maintain a stable static equilibrium
configuration.  Dipole-dipole interactions cause chaotic behavior that
disorders their phases, and hence suppresses SR.  SR arrises from a competition
between these two opposing effects.  This conclusion is inferred from the
theory of non-uniform broadening of spectral lines for
lasers~\cite{friedberg-PLA-40-365,friedberg-PRA-10-1728}.  Consider now a
nonlinear CMS \cite{il,men}, i.e.\ a system
of classical, charged anharmonic oscillators.  Maxwell's equations describe the
electromagnetic field.  Next, assume that there are sufficient oscillators
($N\gg1$), and they occupy a small spatial region of length $L$ such that $l\ll
L\ll\lambda$, where $l=n^{1/3}$ is the characteristic distance between atoms,
and $\lambda$ is the wavelength of the radiation.  Each charge has magnitude
$e$ and mass $m$, and is located on the ends of springs with stiffness
coefficient $k$, at coordinates $\bm{r}_{a}+\bm{\xi}_{a} (a=1,2,...,N)$, fixed
in points $\bm{r}_{a}$, where there are also compensating charges $-e$.  The
equation of motion for the oscillators then takes the form \cite{Landau1975}

\begin{equation}\label{eq:1}
\ddot{\xi_{a}}+\omega_{0}^{2}(1+\gamma\xi_{a}^{2})\xi_{a}=-\frac{2e^{2}
\omega_{0}^{2}}{3mc^{3}}\sum_{b}\dot{\xi_{b}}{}+\frac{e^{2}}{m}\sum_{b\neq a} \nabla_{a}\times\biggl(\nabla_{a}\times\frac
{\xi_{b}(t_{ab})}{r_{ab}}\biggr).
\end{equation}

Here $\nabla_{a}=\partial/\partial\bm{r}_{a}$,
$\bm{r}_{ab}=\bm{r}_{a}-\bm{r}_{b}$, $t_{ab}=t-r_{ab}/c$ represents the
retarded time, $\omega_{0}=\sqrt{k/m}$ is the fundamental frequency of the
oscillators, and $\gamma$ is the nonlinearity parameter.  Substituting the
expression

\begin{equation}\label{eq:2}
\xi_{a}=b[F_{a}(t)\exp(-\imath\omega t)+F_{a}^\ast(t)\exp(\imath\omega t)],
\end{equation}
into equation~(\ref{eq:1})---where $b$ represents the characteristic initial
amplitude of the oscillations---gives

\begin{equation}\label{eq:3}
\dot{F_{a}}+\imath\delta(\vert F_{a}\vert^{2}-1)F_{a}=
\imath\beta\sum_{b\neq a}\nabla_{a}\times{}\biggl[\nabla_{a}\frac{\exp(\imath kr_{ab})}{r_{ab}}\times F_{b}(t)\biggr]
-\frac{1}{2}\beta_{0}\sum_{b}F_{b}.
\end{equation}

In equation~(\ref{eq:3}) the second derivatives of functions $F_{a}(t)$ which vary
slowly in comparison with exponents $\exp(\pm\imath\omega t)$ are omitted, and
a frequency $\omega=\omega_{0}+\delta$, $\delta=3\gamma\omega_{0}b^{2}/2$ is
chosen.  Note, that the case of particles rotating in a magnetic field
(important in a practical sense) corresponds $\delta<0$.  For a small size
system eqution~(\ref{eq:3}) can be rewritten as
 
\begin{equation}\label{eq:4}
\dot{F_{a}}+\imath\delta(\vert F_{a}\vert^{2}-1)F_{a}=
\imath\beta\sum_{b\neq a}\frac{3n_{ab}(n_{ab}F_{b})-F_{b}}{r_{ab}^{3}}-
\frac{1}{2}\beta_{0}\sum_{b}{F_{b}}.
\end{equation}

Where $\bm{n}_{ab} = \bm{r}_{ab}/r_{ab}$, $\beta=e^{2}/(2m\omega_{0})$, and
$\beta_{0}=2e^{2}\omega_{0}^{2}/3mc^{3}$.  The first term on the right hand
side of equation~(\ref{eq:4}) represents the dipole-dipole interaction of the
oscillators, while the second term is analagous to a `viscosity' for the
radiation in the electromagnetic field.  Following Ref.~\cite{il}, we shall
consider one-dimensional oscillators, i.e.\ that dipoles oscillate along the
$\bm{x}$ axis, and consequently, that the vectors $\bm{F}_{a}$ are parallel to
it: $\bm{F}_{a}=\bm{i}F_{a}$, $\bm{i}=(1,0,0)$.  During a given time $t$ we
have $F_{a}(t)=\rho_a(t)\exp(\imath\varphi_a(t))$.  Hence, atoms possess a
dipole moment that is $\bm{d}_a(t)=e\bm{\xi}_{a}=eb\bm{i}\rho_{a}\cos(\omega
t+\varphi_{a})$.

The average radiation intensity of the rapidly oscillating dipoles then is

\begin{equation}\label{eq:5}
I=\frac{e^{2}\omega^{4}b^{2}}{3c^{3}}\sum_{a,b}\vert F_{a}\vert\vert F_{b}\vert\cos(\varphi_{a}-\varphi_{b}).
\end{equation}

Thus, equation~(\ref{eq:4}) represents a system of $N$ oscillators, distributed
arbitrarily, that can be solved by numerical means.  A similar formalism is
described in Ref.~\cite{il}; however, dipole-dipole interactions are neglected.

\section{\label{sec:level2}Results and discussion}

The phasing effect can be described as follows.  Consider a complex plane
$(x,y)=(\Re(F),\Im(F))$ containing $N$ points that each represent the state of
an individual oscillator, where the distance from the origin is simply the
amplitude of oscillation, and the angle is the phase with respect to the
fundamental frequency $\omega_{0}$.  Points with $\omega>\omega_{0}$ rotate
clockwise around the origin; points with $\omega<\omega_{0}$ rotate
anticlockwise.

Initially, the points are placed randomly with equal probability phases on a
circle of unit radius $\rho=1$.  From equation~(\ref{eq:4}), their velocities are
\begin{equation}\label{eq:6}
\bm{v}_{a}=\bm{\omega}(\rho_{a})\times\bm{\rho}_{a}+\bm{f}+\sum_{b}\bm{d}(\rho_{a},\rho_{b};\bm{r}_{a},\bm{r}_{b}).
\end{equation}

Here $\bm{\rho}_a=(\Re(F_{a}), \Im(F_{a}), 0)$,
$\bm{v}_{a}=\dot{\bm{\rho}_{a}}$, $\bm{f}=-\beta_0\sum_{a}\bm{\rho}_{a}/2$, and
$\bm{\omega}(\rho)=(0,0,-\delta(\rho^{2}-1))$,
$\bm{d}(\rho_{a},\rho_{b};\bm{r}_{a}, \bm{r}_{b})$.  The latter dipole-dipole
interaction term is not shown in full for reasons of space.  Note that the
vector $-\bm{f}$ is proportional to the total dipole moment of the system
$\bm{D}=eb\sum_{a}\bm{\rho}_a/2$; and, $\omega(\rho_{a})=0$ at $t=0$.
\begin{figure} 
\centering
\includegraphics[width=10cm]{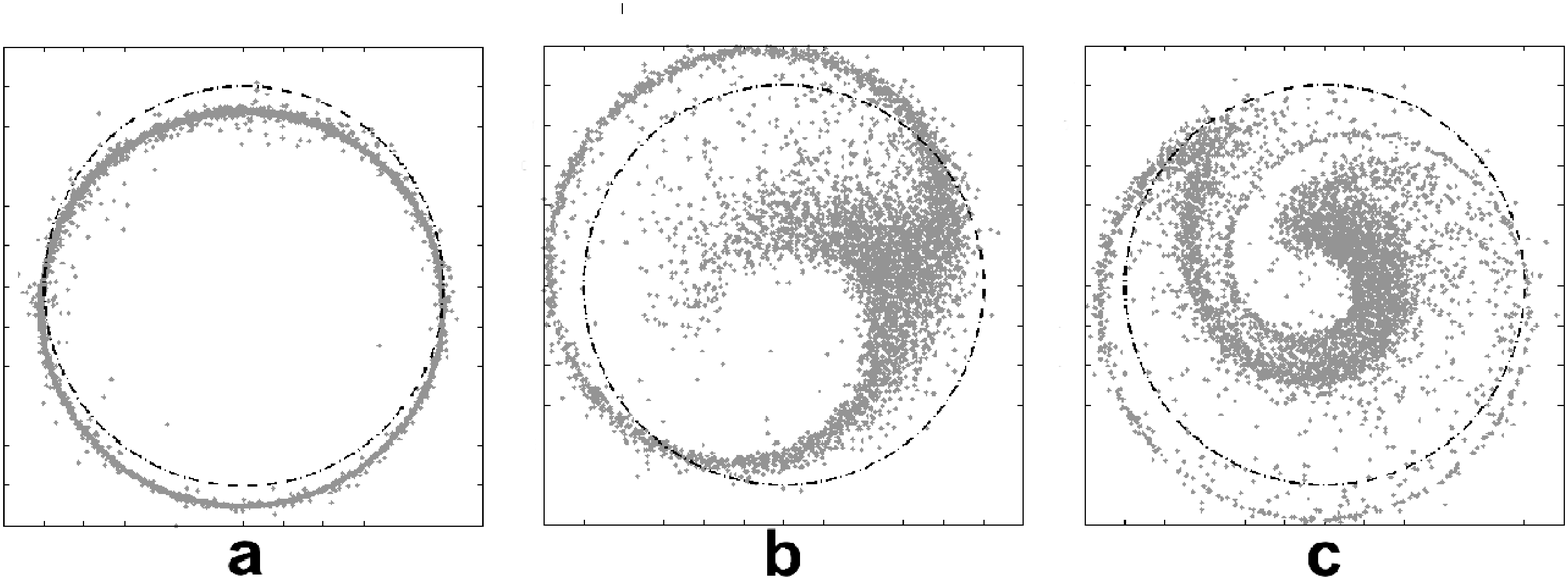}
\caption{\label{fig:f1} Time evolution of the phase distribution of
oscillators.  The dotted line is a circle with unit radius.  The number
of oscillators is $N = 5 \times 10^{3}$. The concentration of oscillators
 $n = 10^{22}$m$^{-3}$ (curve 2 on figure 4).}
\end{figure}
\begin{figure}
\centering \includegraphics[width=10cm]{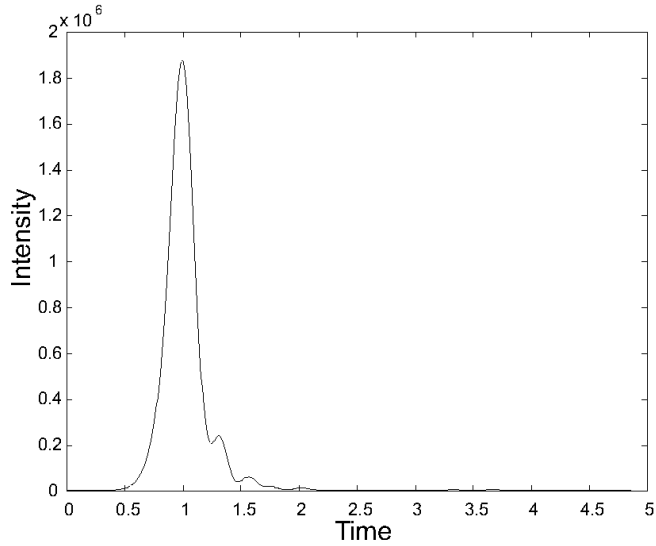}
\caption{\label{fig:f2} Time dependence of the radiation intensity for $N = 5
\times 10^{3}$ (all values in arbitrary units).}
\end{figure}
Notice also that the sign of $\gamma$ affects the direction of rotation
\emph{only}: changing it results in a mirror inversion without any other
consequences.  Points with positive $\gamma$ rotate clockwise outside the unit
circle, and rotate anticlockwise when inside, while the opposite is true when
$\gamma$ is negative.  This symmetry, therefore, is exploited by choosing
$\gamma>0$.

Having established the basis for the model, we next consider how the system
evolves when the density of atoms $n$ is sufficiently small that dipole-dipole
interactions are negligible.  Due to the fluctuations of density distribution
of the oscillators initial phases $\varphi_{a}(0)$, the initial value of the
vector $\bm{f}$ is not precisely zero.  At $t=0$ from equation~(\ref{eq:6}) it
follows that $d\bm{D}/dt = - \bm{D}/\tau_{SR}$, where the characteristic
emission time is
$\tau_{SR}=1/(N\beta_{0})$~\cite{dicke-PR-93-99,gross-PREP-93-301,stenholm-PREP-6-1,men}.

Consequently from equation~(\ref{eq:6}), the system responds by moving 
in a direction opposite to the
dipole moment $\bm{D}$, with a collective net velocity $\bm{f}$.
The system at time $\sim \tau_{SR}$ is displaced a distance 
$\sim \bm{D}(0)/(Ne)$ (see figure~\ref{fig:f1}a).  
The resulting
displacement moves half of the points outside the unit circle ($\rho>1$), and
the other half inside ($\rho>1$).  Hence, points outside the circle will move
in clockwise orbits, while those within circulate the opposite way.  After an
interval $t\sim10\tau_{SR}$, the net motion results in a bunching of points on
the inside of the circle (figure~\ref{fig:f1}b), thus the atoms emit most of
their stored energy in a sharp pulse of coherent radiation (figure~\ref{fig:f2}).
For two-level atoms, the characteristic delay time $t_{0}=\tau_{SR}\log N$
given in ~\cite{dicke-PR-93-99} is consistent with this.  The bunch
subsequently develops into a spiral-shaped distribution (figure~\ref{fig:f1}c).
As it does so, the dipole moment decreases to a minimum, along with the SR
intensity.  The cycle repeats, decaying rapidly (figure~\ref{fig:f2}).
Oscillatory behavior is typical for SR in classical systems of small
size~\cite{il}.  In quantum systems consisting of two-level atoms, SR intensity
oscillations are absent~\cite{dicke-PR-93-99}.

\begin{figure}
\centering
\includegraphics[width=10cm]{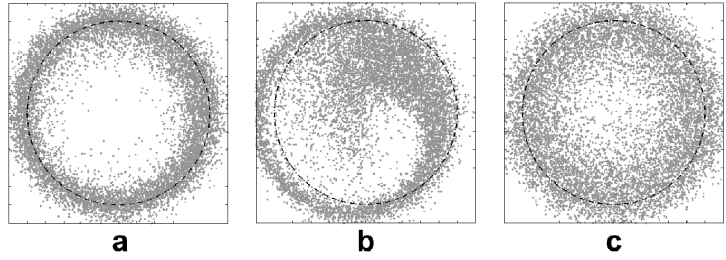}
\caption{\label{fig:f3} Time evolution of the phase distribution of oscillators
in systems with a strong dipole-dipole interaction.  The dotted line is a
circle with unit radius. Figure {\bf a} and {\bf b} corresponds to concentration of 
oscillators $n= 8 \times 10^{22}$m$^{-3}$ (curve 
4 on figure 4 ). Figure  {\bf c} corresponds to $n = 1.8 \times 10^{23}$m$^{-3}$ 
(curve 6 in figure  4).}  
\end{figure}

\begin{figure}
\centering
\includegraphics[width=10cm]{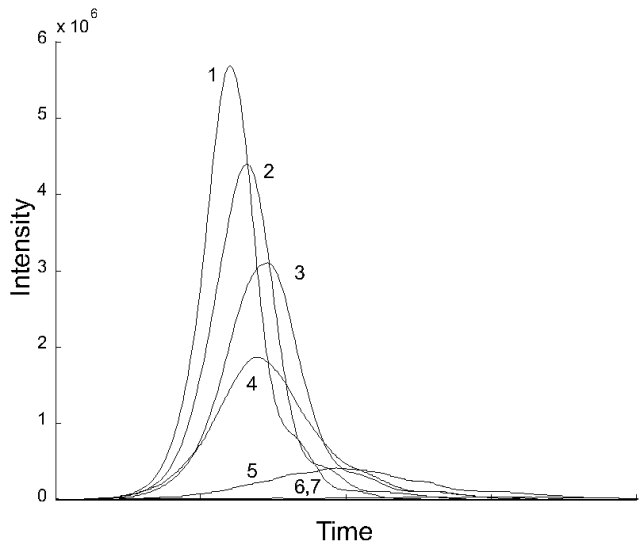}
\caption{\label{fig:f4} Intensity of radiation (arbitrary units) for systems
with different oscillator concentrations $n$ ($10^{22}$m$^{-3}$): $0.083$,
$1.0$, $2.3$, $8.0$, $12.13$, $18.38$, $27.86$, for $1$--$7$ respectively.}
\end{figure}

At high density $n$, dipole-dipole interactions have a significant effect.
Figure~\ref{fig:f3} shows the outcome of equation~(\ref{eq:4}) for large $n$; the
initial conditions are the same as described previously.  Notice that the
points on the phase plane now move in a more chaotic manner than before.  When
$n$ is high, dipole-dipole interactions among adjacent oscillators are strong
and this leads to incoherence.  However, SR is not entirely suppressed.
In spite of the chaotic behavior of dipole-dipole interaction, 
the initial total dipole moment results in bunching of points, and 
correspondently in the SR pulse (figure~\ref{fig:f3}a,b). On figure~\ref{fig:f3}c, where the concentration of 
oscillators was doubled,
dipole-dipole  interaction suppress the bunching.

High density systems
are also complicated by collective effects.  Localized groups of resonant atoms
induce antiphase dipole moments among their neighbors.  This preserves
coherence while screening SR~\cite{men}.
\begin{figure}
\centering
\includegraphics[width=10cm]{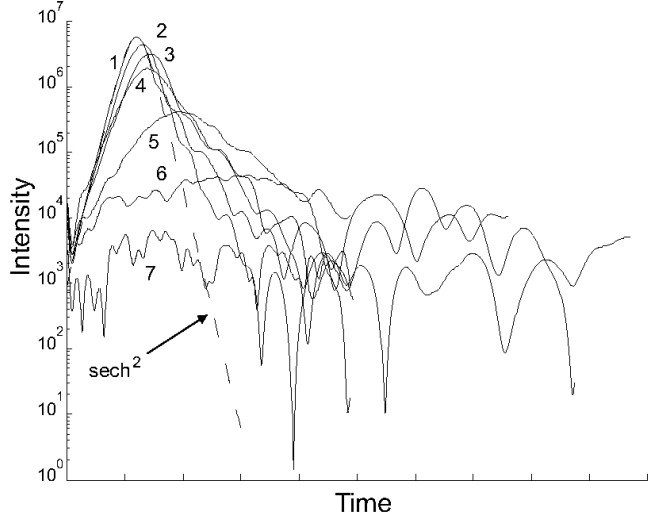}
\caption{\label{fig:f5} Radiation intensity (arbitrary units) versus time
(arbitrary units) for classical systems with different oscillator
concentrations $n$ ($10^{22}m^{-3}$): $0.083$, $1.0$, $2.3$, $8.0$, $12.13$,
$18.38$, $27.86$, for $1$--$7$ respectively.  Case $1$ is compared with the
purely quantum result which varies as $\mathrm{sech}^{2}(t-t_{0})$.}
\end{figure}

\begin{figure}
\centering
\includegraphics[width=10cm]{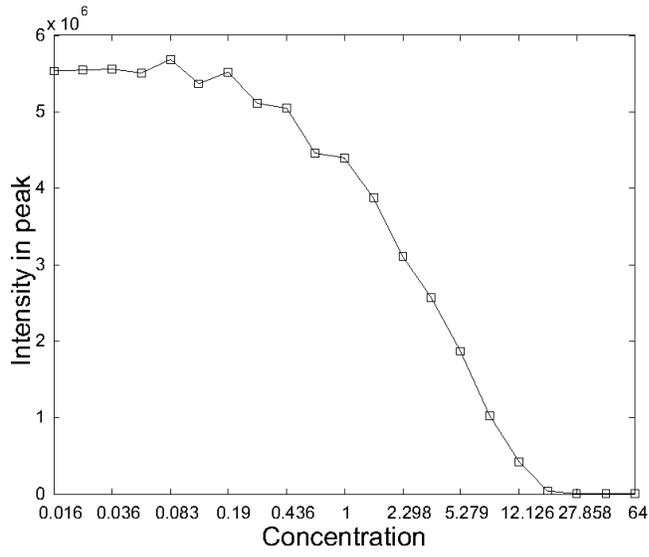}
\caption{\label{fig:f6} Dependence of a maximum of radiation intensity
(arbitrary units) on oscillator density $n$ (in $10^{22}m^{-3}$).}
\end{figure}

The SR delay $t_{0}$ and peak intensity $I_{max}$ also depend on $n$:
increasing $n$ makes $t_{0}$ longer, and $I_{max}$ smaller (see
figures~\ref{fig:f4}, \ref{fig:f5}, and \ref{fig:f6}).  This is a consequence of
the effect of coherence on the collective interactions among the dipoles, which
becomes weaker with increasing $n$.

\begin{figure}
\centering
\includegraphics[width=10cm]{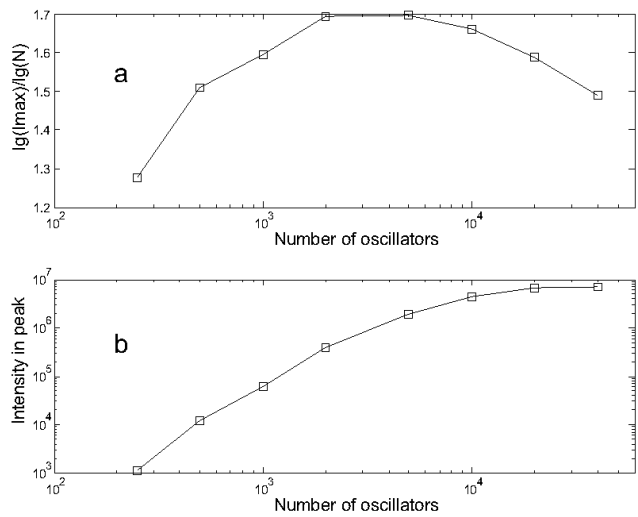}
\caption{\label{fig:f7} Dependencies on the number of oscillators $N$, of (a)
the ratio $\log_{10}(I_{max})/\log_{10}(N)$; and (b) the peak radiation
intensity.}  

\end{figure}

Unlike classical systems, quantum systems do not behave chaotically.  The
intensity varies smoothly with time as described by the following
formula~\cite{dicke-PR-93-99}.
\begin{equation}\label{eq:7}
I(t)=\frac{\hbar\omega_{0}}{4\mu\tau_{N}}(\mu N+1)^{2}\mathrm{sech}^{2}\left(\frac{t-t_{0}}{2\tau_{N}}\right),
\end{equation}
where $\mu$ represents the form-factor of the oscillators' mutual position,
and  $\tau_{N}=1/\beta_{0}$ is the characteristic emission time.  
This curve is plotted
in figure~\ref{fig:f5} to illustrate the difference between the classical and
quantum cases.  When $N$ is large, at $t=t_{0}$, equation~\ref{eq:7} suggests
$I_{max}\sim N^{2}$.  However, the CMS predicts that the exponent
$\alpha=\lg(I_{max})/\lg(N)$ rises to a peak value that is less than two, then
declines as $N$ increases (see figure~\ref{fig:f7}).  Experimental observations
of SR in semiconductors exhibit similar behavior~\cite{zaitsev-SPS-33-1309}.


The results are consistent with Ref.~\cite{men}.  Localized, dynamic metastable
states are formed when the atom density $n$ is sufficiently large.  Each
oscillator perturbs the motion of its nearest neighbors such that their
relative phase differs by $\pi$.  Hence, in effect each oscillator appears to
be screened in a manner analogous to Debye shielding.  This leads to
confinement of electromagnetic fields in the active medium.

\section{\label{sec:level3}Conclusions}

This study examines the phenomenon of superradiation for systems of classical
nonlinear charged oscillators.  The results of our numerical simulations show
that after a characteristic delay time $t_{0}$, a peak in radiated power
occurs, which subsequently decays in a chaotic, oscillatory manner,
superimposed on a $\mathrm{sech}^{2}(t-t_{0})$ background.  SR is also
suppressed progressively with increasing oscillator density $n$.  This behavior
is ultimately a consequence of collective dipole-dipole interactions.  These
both induce incoherence among the oscillators, and cause a screening effect.


Within localized regions,  the individual
dipoles possess correlated moments. Dipoles separated
by sufficient large distances are nearly uncorrelated. As
$n$ increases, the system breaks up into more of these regions.
Each region emits SR impulses independently, resulting in the
chaotic decay described above.

\ack
This work was supported by the Swedish Institute, Lule{\aa} University of 
Technology and by a grant from the administration of Arhangelsk region, Russia: Pomor Young Scientist, 2007, project No. 03-3.

\end{document}